\documentclass[
 reprint,
superscriptaddress,
 amsmath,amssymb,
 aps,
 prx,
]{revtex4-1}

\usepackage{graphicx}
\usepackage{dcolumn}
\usepackage{bm}
\usepackage{enumitem}

\begin{document}

\title{Observation of magnetic-field-sweep-direction-dependent dynamic nuclear polarization under periodic optical electron spin pumping}

\author{Michael Macmahon}
 \affiliation{Department of Physics, University of Michigan, Ann Arbor, Michigan 48109, USA}
 
\author{Joseph R. Iafrate}
\affiliation{Applied Physics Program, University of Michigan, Ann Arbor, Michigan 48109, USA}

\author{Michael J. Dominguez}
\affiliation{Applied Physics Program, University of Michigan, Ann Arbor, Michigan 48109, USA}

\author{Vanessa Sih}
 \email{vsih@umich.edu}
 \affiliation{Department of Physics, University of Michigan, Ann Arbor, Michigan 48109, USA}


\begin{abstract}
Optical pump-probe techniques are used to generate and measure electron spin polarization in a gallium arsenide epilayer in which the electron spin coherence time exceeds the mode-locked laser repetition period. Resonant spin amplification occurs at magnetic fields at which the electron spin polarization excited by successive laser pulses adds constructively. Measurements of Kerr rotation as a function of applied magnetic field and pump-probe time delay reveal nuclear spin polarization that aligns either with or against the external magnetic field depending on whether the applied magnetic field is being decreased or increased. Furthermore, the nuclear spin polarization magnitude varies in proportion to the perpendicular net electron spin polarization as the latter changes due to resonant spin amplification and other causes. We also report an experimental technique that reveals a minutes-long memory of precise field history in the electron-nuclear spin system.
\end{abstract}

\maketitle

\section{\label{sec:intro}Introduction}

The manipulation and hyperpolarization of nuclear spin holds particular interest for potential application in spin-based classical and quantum information processing \cite{Reimer,Urbaszek,Warburton,Chekhovich}. Specifically, understanding how to control and optimize nuclear spin polarization could be applied to generate a hyperpolarization of nuclear spins to enhance magnetic resonance imaging and store information \cite{Reimer}. Furthermore, being able to control nuclear spin polarization would enable pathways for manipulating electron spin polarization and maximizing the electron spin coherence time for electron spin-based information processing \cite{Urbaszek,Warburton,Chekhovich}.

The optical pumping of electron spins has been shown to generate dynamic nuclear polarization in bulk semiconductors \cite{Lampel,KikkawaDNP}, quantum wells \cite{Salis}, and quantum dots \cite{Gammon}. Magneto-optical techniques can monitor this nuclear polarization through its effect (Overhauser field) on the electron spin system \cite{KikkawaDNP,Salis,Gammon}, including in the regime of resonant spin amplification (RSA) \cite{Zhukov2014}. Recently, periodic Overhauser fields have been observed with Voigt geometry time-resolved Kerr rotation (TRKR) in fluorine-doped ZnSe \cite{Heisterkamp,Zhukov2018}. The dynamic nuclear polarization in Ref.\,\,[\onlinecite{Zhukov2018}] is attributed to electron spins, originally perpendicular to the magnetic field, rotated into the magnetic field direction by the optical Stark effect. However, these experiments did not observe any dependence on the direction of the magnetic field sweep \cite{Zhukov2018}. The optically driven electron-nuclear spin system has also recently revealed interesting feedback effects in quantum dots; for example, the nuclear spin polarization has been found to adjust through a feedback mechanism in response to laser frequency \cite{Xu} and applied microwave magnetic field \cite{Vink}.

In this paper, we report Kerr rotation for electron spins in gallium arsenide (GaAs) that dramatically changes based on the direction the magnetic field is swept. This is evidence of a nuclear spin polarization that strongly depends on whether the applied magnetic field is increasing or decreasing. Furthermore, the magnitude of dynamic nuclear polarization changes in response to changes in the magnitude of the transverse electron spin polarization due to RSA, both periodically with applied magnetic field and over time with the field held constant. This causes measurements of Kerr rotation versus an applied external magnetic field to produce peaks that are drastically warped in opposite directions for increasing versus decreasing external fields. Finally, we present an experimental technique of ``steeping" during a sweep of external magnetic field and show that this produces a striking ``steep echo" effect in the subsequent spin signal.

\begin{figure*}
\includegraphics[width=15cm]{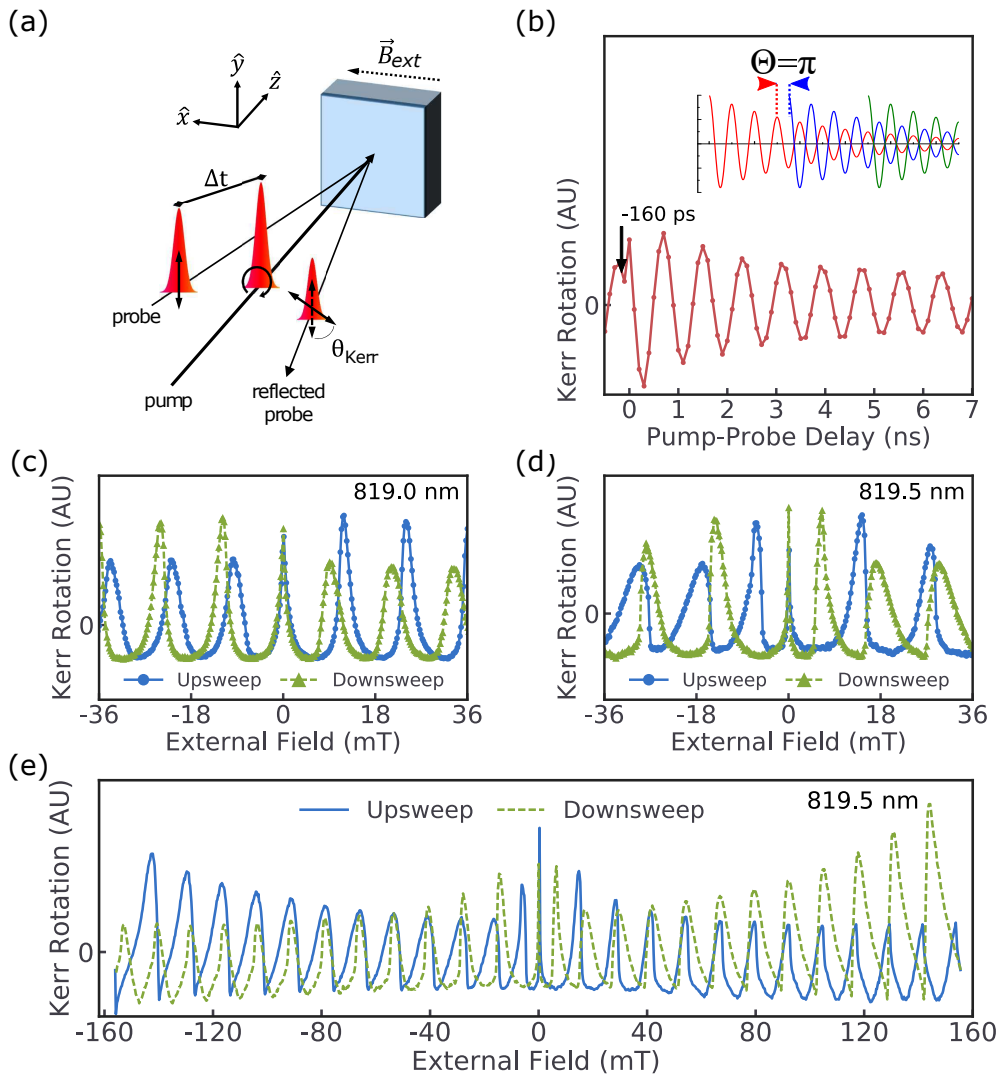}
\caption{\label{fig1}
(a) Experimental geometry for optical pump-probe Kerr rotation measurements.
(b) Time-resolved Kerr rotation measured at a laser wavelength of 818.7 nm with external magnetic field $B_{ext} =$ 200 mT. The arrow at $-160$ ps indicates the pump-probe delay used for magnetic field sweeps. Inset: Due to a spin coherence time longer than the laser repetition period, electron spin packets excited by consecutive laser pulses (denoted with different colors) interfere destructively when $\Theta$, the product of Larmor precession frequency and pump-probe delay (modulo 2$\pi$), equals $\pi$.
(c), (d) Kerr rotation measured as a function of external magnetic field for a fixed pump-probe delay of 13 ns at (c) 819.0 nm and (d) 819.5 nm. The field is swept from $-160$ to $+160$ mT for the upsweep and from $+160$ to $-160$ mT for the downsweep. For clarity, only data from $-36$ to $+36$ mT are shown. Excepting the peak at zero external field, upsweep peaks demonstrate a shift with respect to downsweep peaks. At certain laser wavelengths such as 819.5 nm in (d), the peaks take on a distinctly asymmetric profile absent in (c).
(e) The full upsweep and downsweep excerpted in (d).
}
\end{figure*}

\begin{figure}
\includegraphics[width=8.6cm]{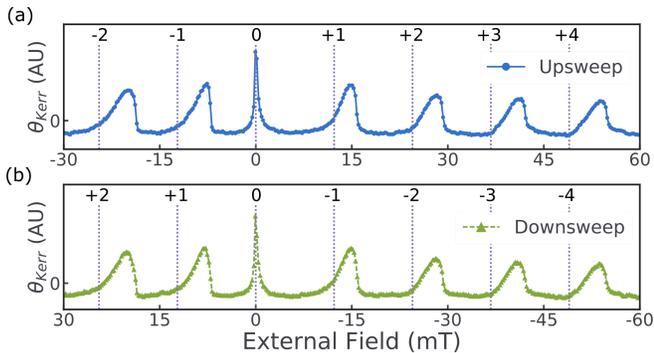}
\caption{\label{fig2}
Kerr rotation measured as a function of external magnetic field for a fixed pump-probe delay of 13 ns. The field is swept from (a) $-80$ to $+80$ mT and (b) $+80$ to $-80$ mT, but for clarity only part of this interval is shown. The peaks are labeled with respect to the peak at zero applied field. By plotting the two peaks encountered before zero and four peaks encountered after for both an (a) upsweep and (b) downsweep, we observe that the field sweep direction, and not sign, determines the shapes of the peaks. Note that the field axis in (b) is reversed. The dotted vertical lines indicate the expected positions of the RSA peaks in the absence of DNP.
}
\end{figure}

\begin{figure*}
\includegraphics[width=14.5cm]{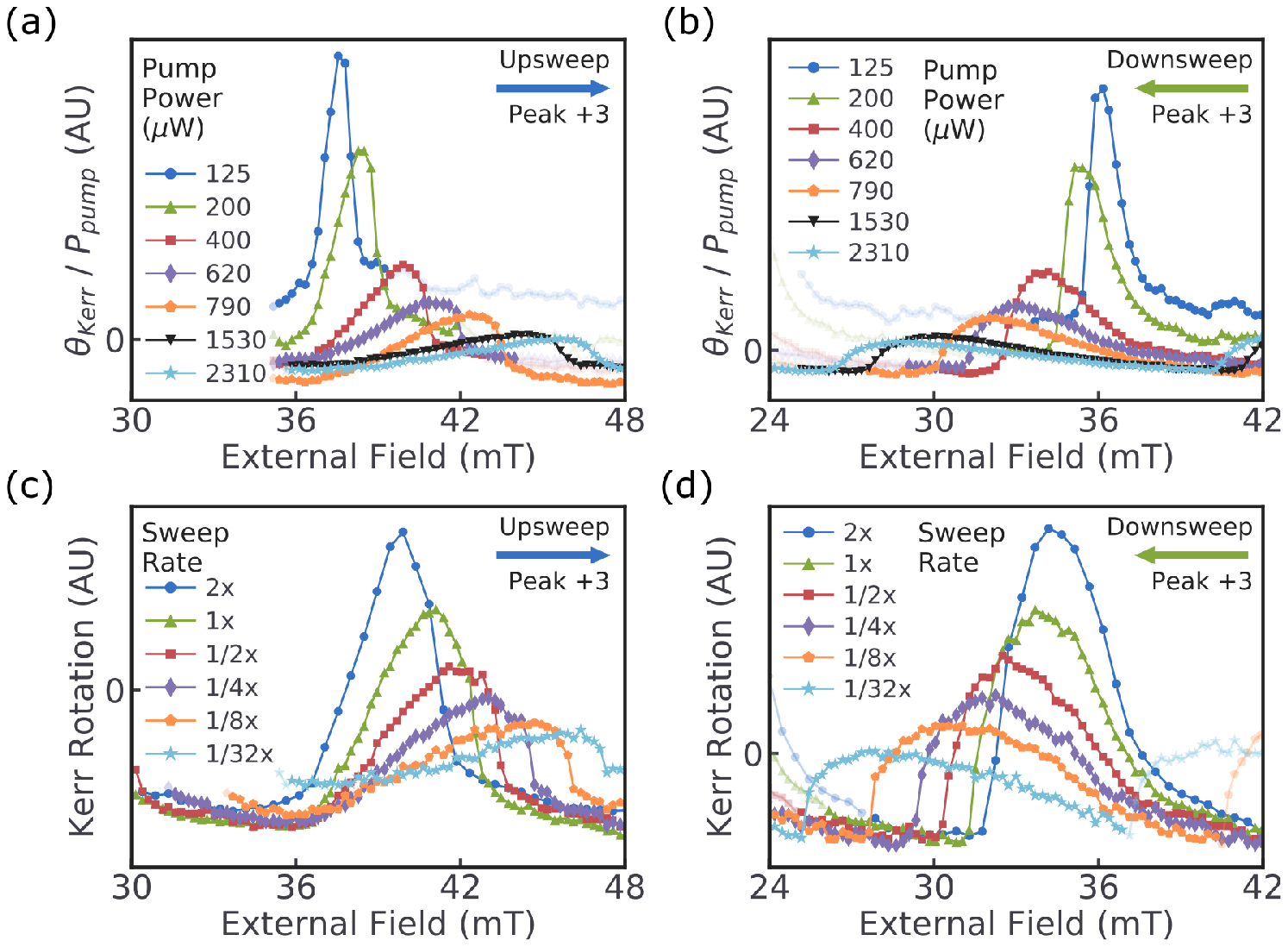}
\caption{\label{fig3}
(a), (b) Peak +3 from Fig. 2 resolved at several pump powers. The Kerr rotation values are normalized by pump power for comparison. As pump power increases, the peak deforms from the standard RSA shape. Both (a) upsweep and (b) downsweep are shown.
(c), (d) The same peak resolved at various sweep rates at a pump power or 530 $\mu$W. Sweep rates are listed as multiples of the default sweep speed of 0.23 mT/s. Here $1\times$ corresponds to the field step timing and spacing used elsewhere in this paper, and the rate was adjusted by varying the duration of field steps. The exception was $2\times$, which used the same timing as $1\times$ but skipped every other field step.
}
\end{figure*}

\section{\label{sec:methods}Methods}

The measurements were performed on a 2-$\mu$m-thick Si-doped GaAs epilayer (doping density $n = 3 \times 10^{16}$ cm$^{-3}$) which was grown on top of a 1-$\mu$m-thick undoped AlGaAs epilayer, grown by molecular-beam epitaxy on an undoped (001) GaAs substrate. We mounted the sample in a helium flow cryostat to maintain the sample temperature at 10 K.

We optically generated and measured electron spin polarization via TRKR \cite{Crooker}. A mode-locked Ti:S laser with a repetition rate of 76 MHz, tuned near the band edge of GaAs, provided both the spin generation (pump) and measurement (probe) beams. A photoelastic modulator alternated the helicity of the pump beam between left- and right-circular polarization at 50 kHz. The linearly polarized probe beam was chopped at 1.37 kHz and reflected off the sample before being split by a Wollaston prism and collected by two optical fibers (A and B) feeding into a photodiode bridge. The prism split the horizontal and vertical components of linear polarization, and a half-wave plate before the prism balanced the proportion of light entering fibers A and B in the absence of Kerr rotation. The photodiode bridge output of $\textrm{A} - \textrm{B}$ corresponds to a direct measurement of Kerr rotation, and lock-in detection was used to filter signal modulated at the frequencies of the photoelastic modulator and optical chopper.

Figure 1(a) depicts the experimental geometry. Unless otherwise noted, measurements are performed with a laser wavelength of 819.5 nm and incident pump and probe powers of approximately 600 and 100 $\mu$W, respectively, measured before the cryostat. A mechanical delay line was used to vary the relative optical path length of the pump and probe beams for TRKR measurements. An electromagnet provided an external magnetic field $B_{ext}$ perpendicular to the optical axis. Electron spins precess about $B_{ext}$ at the Larmor spin precession frequency $g \mu_{B} B_{ext} / \hbar$, resulting in the oscillations seen in Fig. 1(b).

To measure RSA in our sample, we set our delay line to achieve a 13-ns pump-probe temporal separation, equivalent to $-160$ ps in Fig. 1(b). We measured Kerr rotation every 1.1 s, incrementing the applied magnetic field by 0.25 mT every measurement. This allowed us to measure, as a function of these sweeps of the external field, the net spin polarization 13 ns after each pump pulse reaches the sample. To rule out hysteresis in our magnetic field as a cause of our results, we used a gaussmeter to measure the magnetic field at the sample location as the magnet underwent the same sequence and timing of field values as those used to collect the data, and the true, measured field is used in all figures.

\section{\label{sec:results}Results}

\begin{figure*}
\includegraphics[width=15cm]{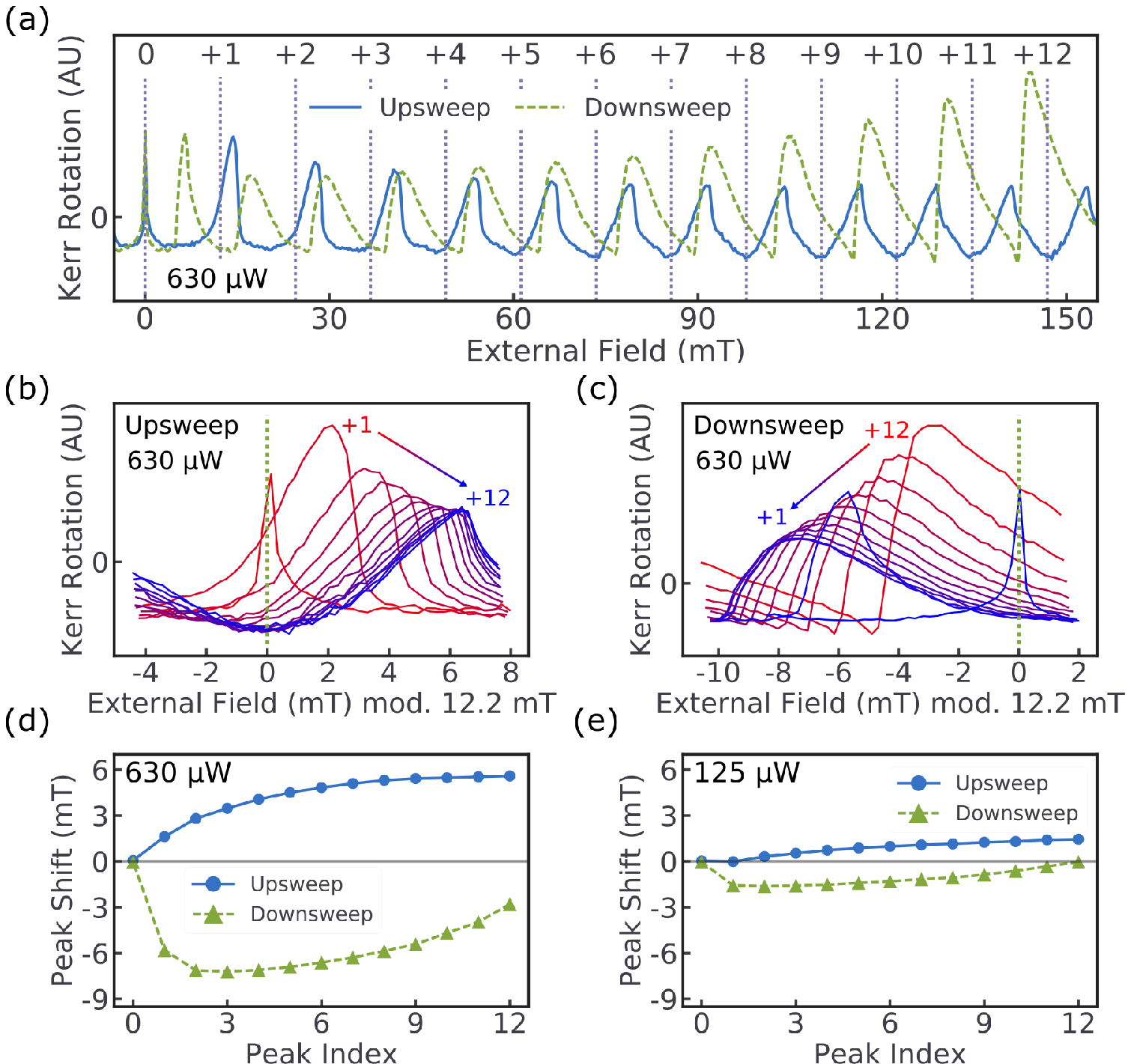}
\caption{\label{fig4}
(a) Kerr rotation measured as a function of external magnetic field for a fixed pump-probe delay of 13 ns. The field is swept from $-160$ to $+160$ mT for the upsweep and from $+160$ to $-160$ mT for the downsweep, but only the field range 0 to $+160$ mT is shown. The peaks are numbered with respect to the peak at zero applied field, as in Fig. 2. The dotted vertical lines indicate the expected positions of the RSA peaks in the absence of DNP.
(b), (c) Peaks $+1$ through $+12$ plotted together as a function of external magnetic field modulo the expected peak spacing of 12.2 mT for field (b) upsweep and (c) downsweep. The dotted vertical line indicates the expected position of the RSA peaks in the absence of DNP. As both field sweeps progress, each successive RSA peak becomes more warped and shifts farther away from the expected position.  
(d), (e) Corrected RSA peak shift as a function of peak index for incident pump power of (d) 630 $\mu$W and (e) 125 $\mu$W. These shifts serve as a measurement of the Overhauser field $B_N$ at each peak location and demonstrate a symmetry between increasing and decreasing external magnetic field. The peak shift is more pronounced under higher incident pump power.
}
\end{figure*}

\begin{figure}
\includegraphics[width=8.6cm]{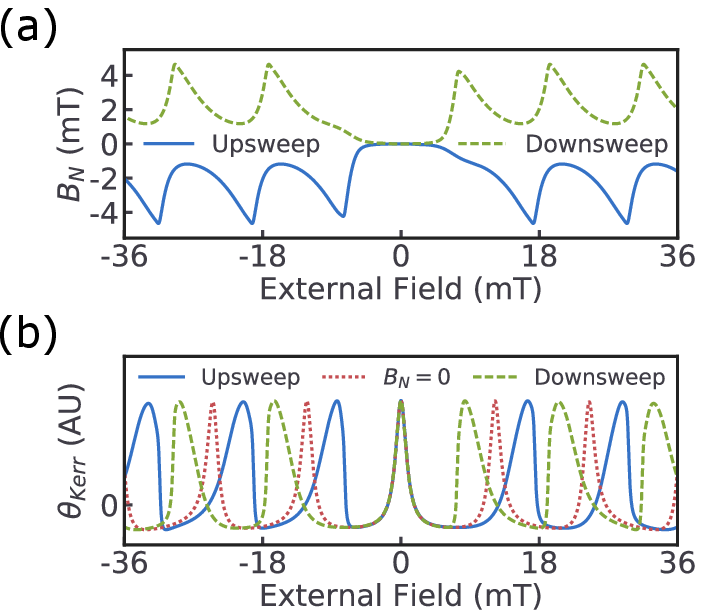}
\caption{\label{fig5}
(a) Simulated Overhauser field $B_N$ (in mT) due to nuclear polarization as a function of external magnetic field and field sweep direction. The field attempts to maintain a magnitude proportional to the total electron spin polarization, continuously approaching said magnitude on a timescale of $T_N =$ 3 s. The simulation sweeps the external field from $-80$ to $+80$ mT, but for clarity only part of this interval is shown.
(b) Simulated Kerr rotation corresponding to the results in (a), demonstrating the peak warping effect. The case of $B_N =$ 0 (no DNP) is shown by the dotted red line.
}
\end{figure}

As shown in Fig. 1(b), a significant Kerr rotation is present at negative pump-probe delay times, indicating a spin coherence time exceeding $T_{rep} =$ 13.16 ns, the repetition period of the laser. As a result, the electron spins generated by each laser pulse will constructively or destructively interfere with the remnants of those from previous pulses depending on $\Theta$, the product of Larmor frequency and $T_{rep}$ modulo 2$\pi$. This phenomenon, resonant spin amplification, can be used to monitor the electron spin dynamics \cite{KikkawaRSA,Trowbridge}.

For most wavelengths and incident pump powers, the resulting pattern of Kerr rotation during a sweep of applied external magnetic field, measured at a fixed pump-probe time delay, consists of sharp peaks separated by $\Delta B_{ext} = h / (g \mu_{B} T_{rep})$ \cite{KikkawaRSA}. Peaks occur when the existing spin polarization is in phase with the polarization induced by the laser pulse each time it illuminates the sample, corresponding to $\Theta = 0$ in the inset of Fig. 1(b). The peak spacing and peak shape are generally identical when sweeping ``up" (negative to positive magnetic field) and ``down" (positive to negative magnetic field). However, under certain measurement conditions, the phenomenon shown in Figs. 1(d) and 1(e) occurs, in which all RSA peaks, except for the central peak at $B_{ext} = 0$, take on a distinctly asymmetric profile with a corresponding peak shift. Sweep direction, sweep speed, field history, and pump power all strongly affect the character of this ``peak warping". As we will demonstrate, this effect is best explained by the presence of a dynamic nuclear polarization (DNP) that rises and falls in proportion to the magnitude of electron spin polarization and reverses for increasing versus decreasing external magnetic fields.

Figure 2 shows the effects of changing magnetic field polarity and sweep direction. When the applied magnetic field is increasing in magnitude, the magnetic field experienced by the electrons is weaker than the one applied, causing each peak to shift to a stronger external field. When the applied magnetic field is decreasing in magnitude, peaks instead shift to a weaker external field, such that in both cases the peaks move away from the current external field. The equivalence of Figs. 2(a) and 2(b) confirms that these differences only depend on whether the external field increases or decreases in magnitude, not upon which direction the field points. This contrasts with prior work demonstrating DNP hysteresis in GaAs/AlGaAs quantum wells, where significant hysteresis occurs only for one polarity of the external magnetic field \cite{Ohno2003, Korenev1992}. Also, the effect described in Ref.\,\,[\onlinecite{Korenev1992}] is attributed to the bistability of the electron-nuclear spin system in the presence of an anisotropic \textit{g} factor in the quantum well, whereas our measurements are conducted on a bulk epilayer. Furthermore, Fig. 2 rules out the possibility that the observed DNP is proportional to the component of time-averaged electron spin polarization parallel to $B_{ext}$ that arises due to a slight projection of the laser along the external field. In this scenario, any effects generated by DNP on the experiments shown in Fig. 2 would depend on external field polarity.

In order to gain a qualitative understanding of this phenomenon, we will now present a phenomenological model. In RSA experiments with our experimental geometry and in the absence of DNP, the following equations adapted from Ref.\,\,[\onlinecite{Trowbridge}] effectively model the time-averaged electron spin polarization component along the laser axis and thus the measured Kerr rotation:
\begin{multline}
    \theta_{Kerr}(B_{ext},T_{delay}) \propto s_z(B_{ext},T_{delay})\\
    = s_r(B_{ext}) \cos \left[ s_\varphi(B_{ext},T_{delay})\right] \text{exp}(-T_{delay}/T_e),
\end{multline}
\begin{align}
    \nonumber s_r(B_{ext}) = &s_0 \{ 1 - 2 \cos \left[ \Theta(B_{ext}) \right] \text{exp}(-T_{rep}/T_e)\\
    &+ \text{exp}(- 2 \text{ }T_{rep}/T_e) \}^{-1/2},
\end{align}
\begin{multline}
    s_\varphi(B_{ext},T_{delay}) = (g \mu_{B} T_{delay} / \hbar) B_{ext}\\
    + \tan^{-1}\left(\frac{ \text{exp}(- T_{rep}/T_e) \sin \left[ \Theta(B_{ext}) \right]}{1 - \text{exp}(- T_{rep}/T_e) \cos \left[ \Theta(B_{ext}) \right]}\right),
\end{multline}
\begin{equation}
    \Theta(B_{ext}) \equiv (g \mu_{B} T_{rep} / \hbar) B_{ext} \pmod{ 2 \pi },
\end{equation}
%
%
where $T_{e}$ is the electron spin lifetime, $s_r$ is the spin polarization magnitude at zero pump-probe delay ($T_{delay} = 0$), and $s_z$ is the spin polarization along the optical axis $\hat{z}$. To reconcile Eq. (1) with the peak warping we have discussed, we introduce an Overhauser field $B_N$ parallel to the external field $B_{ext}$ that also acts on the electron spin system. Then, we substitute $B_{ext}$ with $B_{ext} + B_{N}$ in Eqs. (1)-(4) and allow this new field to vary with a timescale $T_N$. Under these conditions, $B_N$ can be thought to locally shift, shrink, and stretch $\theta_{Kerr}(B_{ext})$ and, when properly chosen, reproduce the qualitative features of the phenomenon shown in Figs. 1 and 2.

In our phenomenological model, we propose that $B_N$ obeys three heuristic properties that qualitatively reproduce the peak warping seen in our experimental results. The properties are as follows:

\begin{enumerate}[label=(\arabic*)]
    \item $B_N$ attempts to maintain a magnitude proportional to $s_r$ as described in Eq. (2). This change is limited by the nuclear polarization timescale $T_N$.
    \item $B_N$ is zero in the vicinity of $B_{ext} =$ 0 and the optical nuclear magnetic resonances of spin-polarized nuclear species.
    \item $B_N$ is aligned antiparallel to the external field when the external field magnitude increases and parallel when the external field magnitude decreases.
\end{enumerate}

The magnitude of nuclear polarization dynamically rises and falls in response to changes in the magnitude of the electron spin polarization, but unlike the electron spins, the nuclear spins align collinear to the external magnetic field. This is property 1. Figures 3(a) and 3(b) provide evidence that the magnitude of DNP is proportional to the magnitude of electron spin polarization. Each RSA peak occurs at a field satisfying $\Theta(B_{ext} + B_N) =$ 0 [see Eq. (4)]. As a result, the shift of each RSA peak in a sweep of $B_{ext}$ is a measurement of $B_N$, assuming the \textit{g} factor of the spins is constant (note that a change in \textit{g} factor would lead to a change in peak spacing that we do not observe). Because the magnitude of electron spin polarization is proportional to the power of the pump laser, we can interpret Figs. 3(a) and 3(b) as indicating a correlation between nuclear polarization and the magnitude of electron spin polarization.

When the external magnetic field approaches zero, nuclear polarization is lost due to nuclear dipole-dipole interactions, so our model must suppress $B_N$ around $B_{ext} = 0$, hence the first part of property 2 \cite{Meier}. The need for the second part of property 2 manifests dramatically in the data shown in Figs. 4(a) and 4(c), where peak +1 for the downsweep cuts off prematurely around 6 mT. This is because, despite Larmor precession, there is a nonzero time-averaged electron spin polarization that produces a Knight field which oscillates in direction and magnitude at the photoelastic modulator frequency (50 kHz). This results in optical nuclear magnetic resonance (optical NMR) in the vicinity of 6 mT \cite{Heisterkamp, Kawakami}. When $B_{ext} + B_N$ reaches a resonance, the nuclear polarization is lowered or eliminated, and the RSA peak shifts away from the current external field, hence the peak cutoff.

Property 3 has no obvious theoretical basis that we know of, but the reversal of peak warping and peak shifts upon reversal of sweep direction is unambiguous. Figures 3(c) and 3(d) show that these reversals are not artifacts due to sweeping the magnetic field much faster than the timescale of DNP, which could cause transient effects to manifest differently for upsweeps and downsweeps. Instead, asymmetry between upsweeps and downsweeps becomes even more pronounced as the system spends more time in the feedback loop between changes in DNP and RSA which we will now describe.

Figure 5 shows the output of a simulation that exhibits peak warping behavior using the model heuristics with $T_N =$ 3 s. This demonstrates how the peak warping arises naturally from these heuristics. To understand this, consider what happens during an external field sweep. As $B_{ext}$ approaches a RSA peak from either direction, electron spin polarization rises, followed by $B_N$. The change in net field shifts the peak away, but this movement is limited by the need to maintain enough spin polarization to sustain $B_N$. When the peak is finally passed, the resulting drop in spin polarization causes a corresponding decrease in $B_N$ that moves the peak back towards its original position and further drops spin polarization, further decreasing $B_N$ and so forth. The result is a sharp drop in signal until the next peak.

Figure 4(a) presents the results of a field sweep upwards from zero field to 160 mT and back. Figures 4(b) and 4(c) superimpose these peaks to allow examination of how each successive peak differs from the last during the upsweep and downsweep, respectively. Figure 4(d) charts the shifts in peak location, and Fig. 4(e) shows these shifts for a lower pump power. Note that RSA itself produces a noticeable warping of high-field peaks in field sweeps \cite{KikkawaRSA}, as can be seen most clearly in the shape of the higher-numbered peaks. This warping effect was isolated and corrected for in Figs. 4(d) and 4(e). As noted earlier, the shift of each peak serves as a  measurement of $B_N$, so Figs. 4(d) and 4(e) measure the trend of $B_N$ across many peaks. Despite the variations in $B_N$ around each peak that produce peak warping, on average $B_N$ steadily rises over the course of the sweep, with a much stronger effect at higher pump power (consistent with property 1).

\begin{figure*}
\includegraphics[width=15cm]{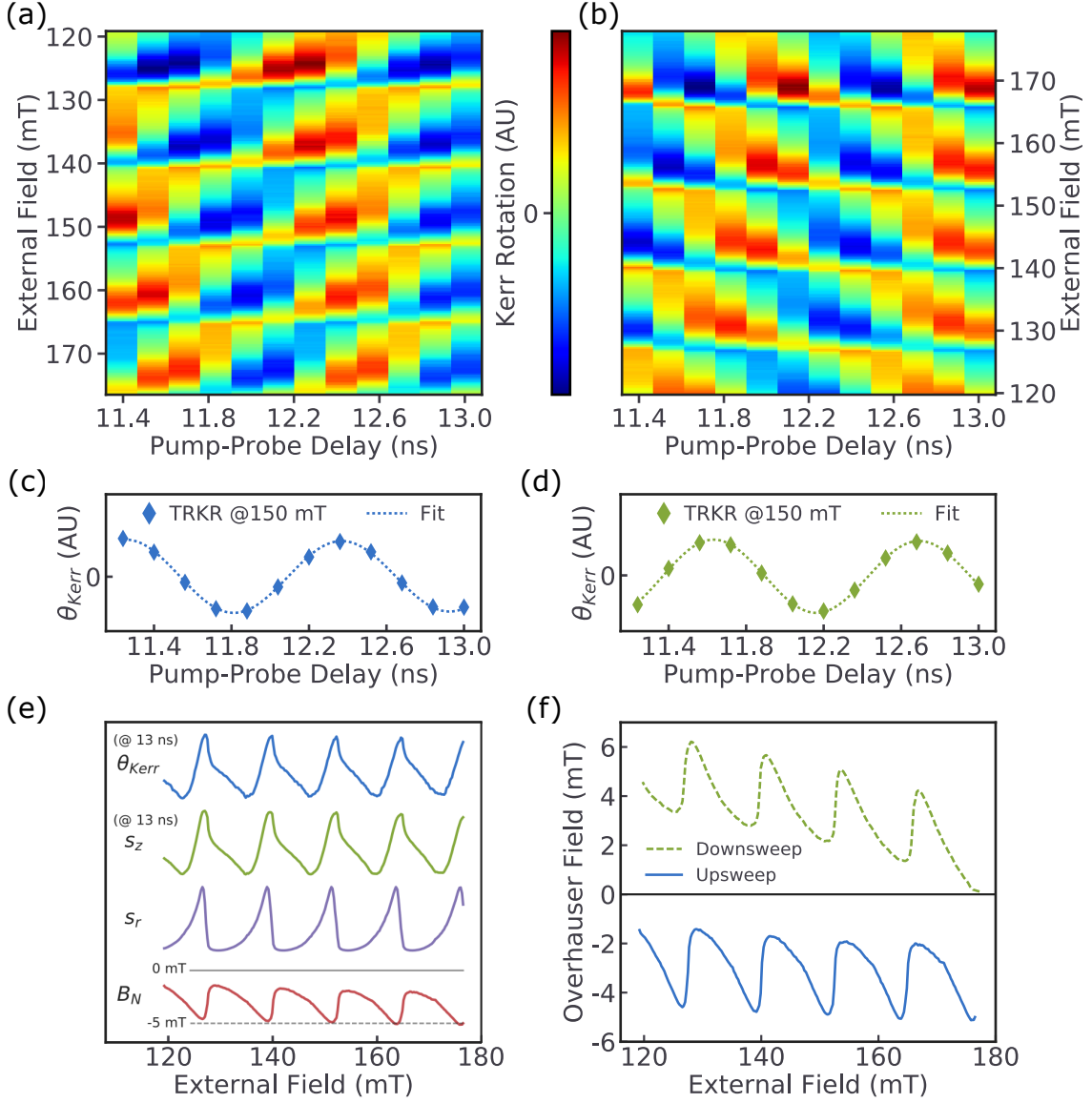}
\caption{\label{fig6}
(a), (b)  Kerr rotation measured as a function of external field and pump-probe delay for (a) increasing and (b) decreasing external field. The data are compiled by scanning the external field while each pump-probe delay is held constant. Incident pump power is (a) 510 $\mu$W and (b) 540 $\mu$W.
(c), (d)  Equivalent time-resolved measurement of Kerr rotation extracted from the panel above at the indicated external field. A fit to Eq. (1) is shown. The large discrepancy in phase despite equivalent external fields demonstrates the difference in DNP between upsweeps and downsweeps.
(e) Kerr rotation $\theta_{Kerr}$ measured as a function of external field for a fixed pump-probe delay of 13 ns, taken from (a). Corresponding parameters extracted from fits of the time-resolved data in (a) are shown below. The spin polarization $s_z$ along the laser axis at 13 ns extracted from the TRKR fits is observed to match the measured $\theta_{Kerr}$. The Overhauser field $B_N$ required to match the observed phase (modulo 12.2 mT) is determined from the measured external field and can be seen to follow the spin polarization magnitude $s_r$. Horizontal lines provide an indication of scale for $B_N$.
(f) Overhauser field $B_N$ extracted from the fits for increasing and decreasing external magnetic field in (a) and (b).
}
\end{figure*}

While RSA peak shifts can be used to measure $B_N$ near the magnetic fields at which peaks occur, measurements as a function of pump-probe time delay can be used to determine $B_N$ at other magnetic fields from changes to the spin precession frequency. However, at low magnetic fields, the spin precession period may be greater than the available delay time range. Figures 6(a) and 6(b) show the results obtained from Kerr rotation measurements as a function of applied magnetic field from 119 to 177 mT at several pump-probe delay times. For each field step, measurements taken at the various delay times constitute TRKR and are fit according to Eq. (1). We show two examples of fitted TRKR in Figs. 6(c) and 6(d); these fits allow us to accurately extract $B_N$. Figures 6(e) and 6(f) demonstrate that $B_N$ has the same periodicity in external field as the RSA peaks. Furthermore, $B_N$ appears to rise and fall with the $s_r$ extracted from the fits. Figure 6(f) shows that $B_N$ differs in sign for increasing and decreasing magnetic fields. These findings provide strong backing to properties 1 and 3 of our model and are consistent with the explanation for peak warping we have described.

Figures 4(b)-4(e), 6(e), and 6(f) all show a slow buildup of DNP over several RSA periods. It should be noted that the simulation in Fig. 5 predicts no such long-term trend, and using a much longer $T_N$ neither produces significant peak warping nor effectively matches the overall trajectory of DNP seen in these figures. This is why the model describes the properties as ``heuristics"; we do not know how to marry the short-timescale peak warping behavior with the slow buildup of DNP over many RSA periods with a single timescale $T_N$. We will now examine these trends in more detail.

Figures 4(d) and 4(e) suggest the DNP buildup requires several RSA periods to level off, explaining why the effect does not diminish over the small field range in Fig. 6. Comparing Fig. 4(b) with Fig. 4(c), we see that despite opposite trends in the warping of peak shape due to RSA, each successive peak measured (indicated by the direction of the arrow) grows shorter and wider and is shifted farther from the no-DNP baseline than the last. This occurs regardless of whether the new peak corresponds to a higher or lower field than the last, as Figs. 4(d) and 4(e) demonstrate. Figures 6(e) and 6(f) explicitly confirm this result: DNP builds up in magnitude slowly over several RSA periods for both increasing and decreasing external fields, although the effect is stronger in the latter for these higher-field results. Note that this means the strength of DNP depends on how much of the sweep has elapsed (the number of peaks passed, the time elapsed during the sweep, etc.) rather than directly depending on the strength of the external field. We suspect this condition rules out many other potential explanations for the data.

For example, Heisterkamp \textit{et al.} explain their results in ZnSe using a model where DNP arises in the vicinity of the external field that produces optical NMR, which occurs in the same manner discussed in our discussion of property 2 \cite{Heisterkamp}. Since electron spin polarization perpendicular to the external field produces nuclear polarization, this model of DNP seems potentially relevant to our results. However, in this model, only the proximity of the current external field to a nuclear species' NMR resonance field affects DNP strength. This is not a problem for explaining their own measurements because they do not report any difference in their results for sweeps of increasing versus decreasing field. In contrast, we find the magnitude of DNP bears little (if any) relationship to the external field's proximity to our optical NMR range of roughly 4-7 mT, and this model explains neither the apparent reversal of DNP based on sweep direction nor the increase in DNP magnitude over multiple RSA periods.

Another potential explanation involves the nuclear spin polarization that arises in a low-temperature nuclear spin system. The Knight field generated by electron spin polarization orthogonal to $B_{ext}$ is expected to produce a cooling of the nuclear spin system and a corresponding $B_N$ parallel to $B_{ext}$ \cite{Meier}. This might produce a $B_N$ that follows $s_r$ as we observe. However, it is not clear how this nuclear cooling model is consistent with property 3, as the DNP we observe changes sign for increasing versus decreasing external fields.

It is important to note that the measurements of peak shift in Figs. 4(d) and 4(e) and of $B_N$ in Figs. 6(e) and 6(f) all have a slight dependence on the precise \textit{g} factor expected in the absence of DNP. Small changes to this value manifest as a roughly linear offset in peak shift/$B_N$ versus external field, potentially as high as $\pm$3 mT at an external field of 180 mT. Different choices of wavelength produce subtle changes in apparent \textit{g} factor, eliminating one possibility for a definitive reference measurement unaffected by DNP. Figures 3(a) and 3(b) show that even small pump powers result in DNP at the wavelength used, obfuscating the \textit{g} factor derived from RSA peak spacing. In the absence of a reliable measurement of \textit{g} factor under experimental conditions but under the assumption that $B_N$ initially starts small, we choose a baseline \textit{g} factor that minimizes the initial (high external field) downsweep $B_N$ in Fig. 6(f). This choice results in the 12.2-mT expected RSA periodicity. A slightly different baseline \textit{g} factor for 125-$\mu$W pump power is used to calculate the peak shifts in Fig. 4(e), resulting in a periodicity of 12.1 mT. This value minimizes the magnitude of downsweep $B_N$ at peak $+12$.

\begin{figure}
\includegraphics[width=7.5cm]{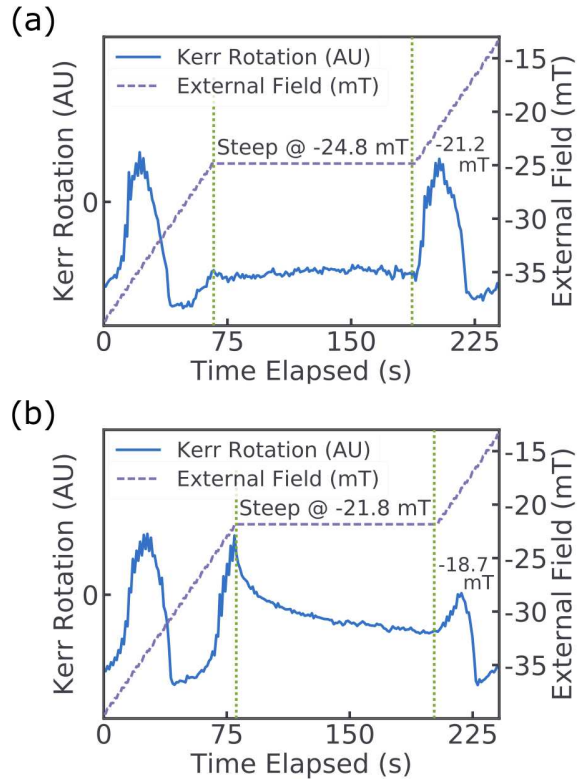}
\caption{\label{fig7}
Kerr rotation (solid line) measured as a function of time elapsed during a field sweep for a fixed pump-probe delay of 13 ns and laser wavelength of 819.3 nm. The external magnetic field (dotted line) is swept up from $-80$ mT, incremented at a rate of about 0.23 mT/s, to the desired ``steep" field and held constant for 2 min. The field is then swept up to $+25$ mT, although only a portion of each field sweep is shown here for clarity. Data are shown for steep fields (a) in the trough between RSA peaks and (b) on the rising edge of a RSA peak. The peak location changes as a result of steeping, as shown by the labeled peaks.
}
\end{figure}

\begin{figure*}
\includegraphics[width=17cm]{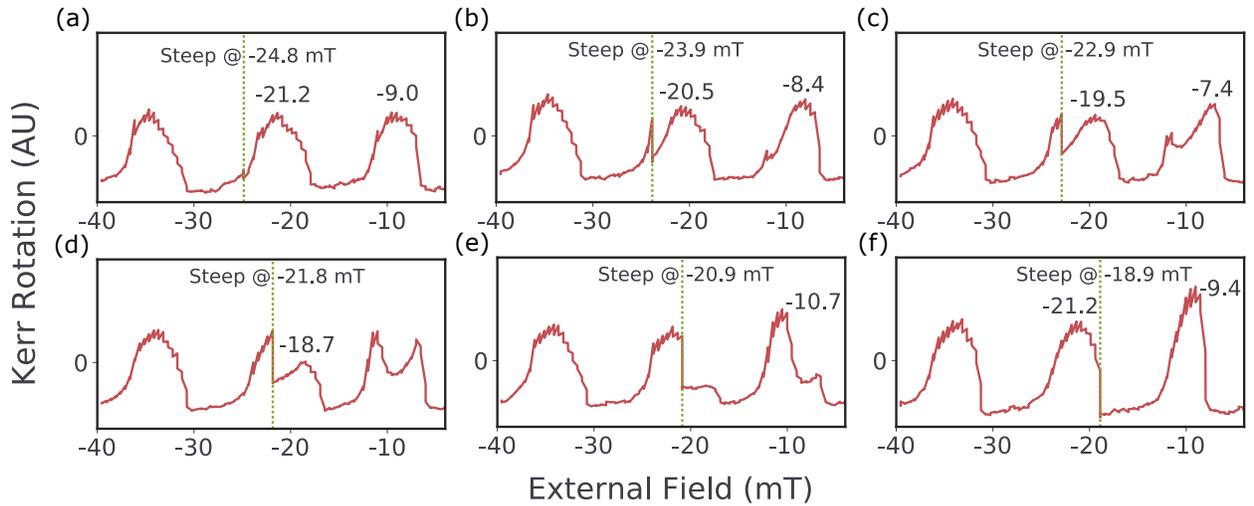}
\caption{\label{fig8}
Kerr rotation measured as a function of external magnetic field for a fixed pump-probe delay of 13 ns and laser wavelength of 819.3 nm. The external magnetic field is swept up from $-80$ mT, incremented at a rate of about 0.23 mT/s, to the desired steep field and held constant for 2 min. The field is then swept up to $+25$ mT, although only peaks $-3$, $-2$, and $-1$ are shown. The chosen steep field influences the locations of peak $-2$, as shown by the labels, with steep fields close to the maximum of peak $-2$ yielding the greatest shift. Steep field (a) $-24.8$ mT is far enough from the center of peak $-2$ to not cause that peak to shift, as is evident by comparison to (f) $-18.9$ mT. Furthermore, peak $-1$ is heavily deformed, in accordance with the chosen steep field, so as to ``echo" the shape of peak $-2$.
}
\end{figure*}

\begin{figure*}
\includegraphics[width=14.5cm]{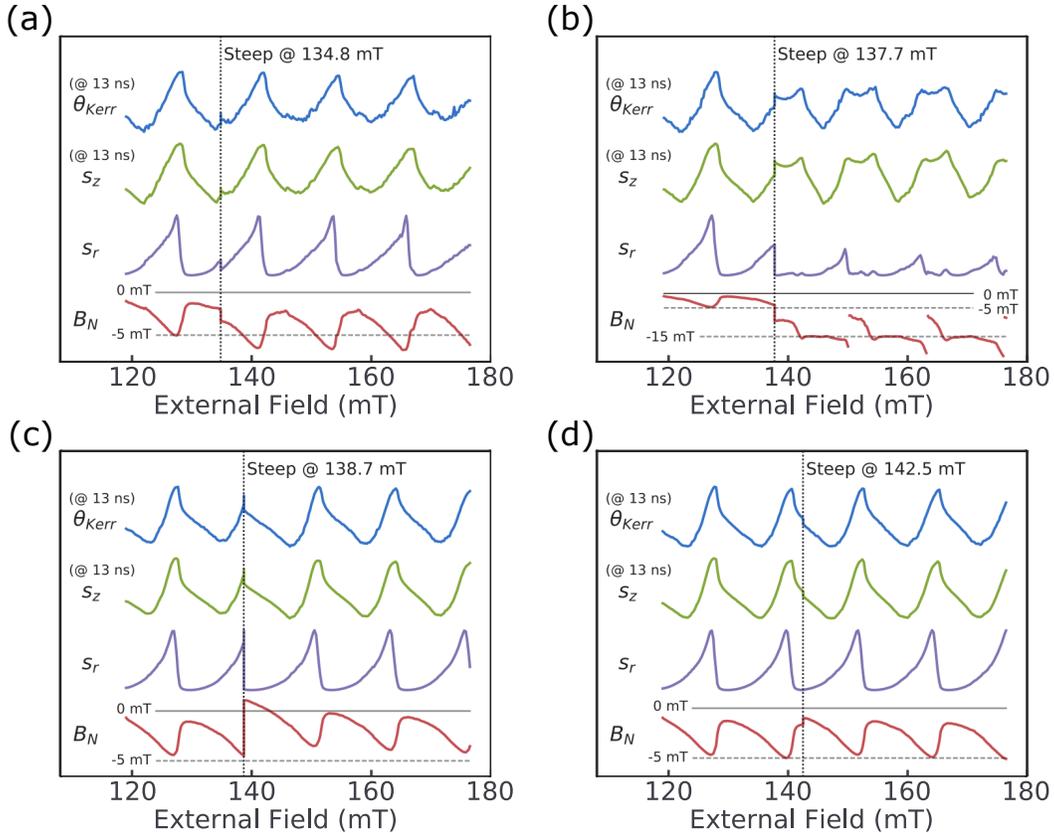}
\caption{\label{fig9}
Kerr rotation measured as a function of external field for a pump-probe delay of 13 ns, at incident pump powers of (a) 660 $\mu$W, (b) 700 $\mu$W, (c) 620 $\mu$W, and (d) 640 $\mu$W. The external field and delay ranges are identical to those covered in Fig. 6, and again, corresponding parameters extracted from fits of the time-resolved data are shown below. However, a 2-min pause (``steep") has been implemented at (a) 134.8 mT, (b) 137.7 mT, (c) 138.7 mT, and (d) 142.5 mT, as in Fig. 7. Horizontal lines provide an indication of scale for $B_N$. The $B_N$ in (b) exhibits three discontinuities of 12.2 mT, as discussed in the text.
}
\end{figure*}

The experiments shown in Figs. 7, 8, and 9 further support our model of peak warping and the relationship between $B_N$ and $s_r$. In each experiment, the magnetic field sweep is paused for 2 min at various external field values. We measure the Kerr rotation while the external field is stationary (the steep period) before resuming the sweep. In Fig. 7(b) we steep on the rising edge of a RSA peak, and the Kerr rotation falls on a timescale of 10 to 100 s. Before the steep, the electron spin system had just emerged from its minimal polarization in the RSA trough, which was coupled with a corresponding drop in $B_N$. This increase in electron spin polarization now causes $B_N$ to begin to steadily rise during the steep. To see why this causes the observed drop in Kerr rotation, we will return to our earlier explanation of peak warping. During the steep the external field is static, and the peak slowly moves away, causing the observed drop in Kerr rotation. The Kerr rotation approaches an asymptotic limit as the peak moves far enough away that the electron spin polarization is just strong enough to sustain its corresponding steady-state $B_N$. In contrast, Fig. 7(a) shows that when steeping while the external field is in a RSA trough, Kerr rotation actually rises slightly as $B_N$ shrinks even more in the minimal electron spin polarization. The peak after the steep is displaced in proportion to the rise of $B_N$ during the steep, supporting the ``peak shifting away" interpretation for the drop in Kerr rotation.

Figure 8 provides more evidence that $B_N$ indeed rises and falls with the electron spin polarization as described in the last paragraph. Figure 8(f) shows that without any steeping, a peak occurs at $B_{ext} = -21.2$ mT. However, Figs. 8(a)-8(d) show that steeping successively closer to that peak results in the peak being pushed back increasingly farther, to a maximum extra peak shift of $+2.5$ mT when the steep occurs at $-21.8$ mT, right at the cusp of the peak and where electron spin polarization is maximized. This extra peak shift is a measurement of how much $B_N$ rose during the steep, and we see that this can vary between 0 and 2.5 mT based on the electron spin polarization at which the steep occurred. Furthermore, this can be done with the same results on any peak for increasing or decreasing $B_{ext}$, so clearly, the trajectory of $B_N$ varies in a complex way based on the precise external field history of the system down to the millitesla level, providing compelling evidence for property 1.

Figures 8(e) and 8(f) demonstrate the interesting case of steeping on the falling edge of the RSA peak, in which we also see a drop in Kerr rotation. However, unlike in Fig. 8(a)-8(d), the peak is not resolved subsequent to the steep. After $B_N$ has naturally risen to its maximum around the RSA peak due to the large $s_r$, steeping at a smaller $s_r$ on the falling edge instead lowers $B_N$. Since this moves the peak away [although in the opposite direction as in Figs. 8(a)-8(d)], $s_r$ falls further, and the effect is self-reinforcing, manifesting in the precipitous drop of Kerr rotation. The sharp decrease in $B_N$ is also demonstrated in the forward shift of the next peak, but the validity of this latter interpretation is cast into doubt by the other notable feature of Fig. 8: the strange behavior of the peak at $B_{ext} \approx -10$ mT. The precise location of the steep on the previous RSA peak causes a striking change in the shape of this peak, even splitting the peak in two. In this case only one peak remained before $B_N$ was erased in the vicinity of $B_{ext} =$ 0, but in the general case this peak deformation occurs on all peaks subsequent to the steep, albeit to a lesser degree with each successive peak. We do not attempt to explain this steep echo behavior here, but we are currently examining the effect further.

As a further test of the explanations we offer for Figs. 7 and 8, we repeat the experiment shown in Fig. 6(a) with 2-min steeps added at four different external fields. The results of the associated TRKR fits are shown in Figs. 9(a)-9(d) and can be contrasted to their no-steep counterpart in Fig. 6(e). Figure 9(a) confirms our earlier interpretation for Fig. 7(b) of a rising-edge steep causing an increase in $B_N$. Figure 9(d) similarly supports our interpretation for Fig. 7(a) of a trough steep causing a slight decrease in $B_N$. While $s_r$ is seemingly not yet falling at the steep location, in Fig. 9(c) the RSA peak is quickly passed as the steep begins, after which the situation appears to play out according to our interpretation of Figs. 8(e) and 8(f). Furthermore, Fig. 9(c) confirms that the next peak is indeed moved forward, and there is no sign of the potentially complicating steep echo effect. In each case, we are able to extract $s_r$, and it is this that predicts the behavior of the steep, not Kerr rotation. This provides one last item of evidence supporting the causal role of $s_r$. Figure 9(b) demonstrates the strangeness of the steep echo effect. To explain Fig. 9(b) in terms of the experimentally observed TRKR rather than fit results, we find in RSA peaks subsequent to the steep that midpeak TRKR amplitude dips and phase quickly shifts by more than $\pi$. This occurs at external fields corresponding in the RSA cycle to the original steep. As shown in the fit, these results are consistent with a sharp rise in $B_N$ every peak. This rise is not offset by the usual postpeak measurable drop in $B_N$, but we are skeptical that the data represent a large, real, cumulative rise in $B_N$ every peak. However, due to the limited length of our delay line, we cannot unambiguously decouple changes in phase from changes in Larmor precession frequency. To represent this uncertainty, we show a discontinuity in $B_N$ at each steep echo in Fig. 9(b) by an indistinguishable 12.2 mT rather than allow $B_N$ to rise beyond $-20$ mT. Future work is required to fix this experimental limitation and further explore this phenomenon.

\section{\label{sec:conclusion}Conclusion}
We have shown the existence of DNP that varies in magnitude with the net electron spin polarization generated by the resonant amplification or nullification of successive packets of optically generated electron spins. Furthermore, this nuclear polarization is parallel to the external magnetic field and perpendicular to the electron spins upon which it depends. The phenomenological model we present captures the subtle time-dependent behaviors of the nuclear polarization. To the degree this model is accurate it raises the mystery of the apparent difference in nuclear polarization for increasing versus decreasing fields. Finally, we have shown that steeping during a sweep of external magnetic field produces a mysterious steep echo effect in the subsequent spin signal that warrants further investigation.

\begin{acknowledgments}
The authors would like to thank H.-W. Hsu and M. Goyal for their contributions. J.R.I. was supported by the Department of Defense through the National Defense Science and Engineering Graduate Fellowship (NDSEG) program. The work at the University of Michigan is supported by the National Science Foundation under Grant No. DMR-1607779. Sample fabrication was performed at the University of Michigan Lurie Nanofabrication Facility.
\end{acknowledgments}

\end{document}